# Avalanche Prediction and Dynamics using Temperature Variance , Grain Size Variance and Flow Regimes

Aditya Sharma

**Abstract**

We investigate the effects of temperature variance, grain size variation, flow regimes, and the use of Support Vector Machines (SVMs) in avalanche studies. The temperature variance experiments involved ice single crystals and polycrystals, revealing that the scale-free pattern of avalanche sizes remains consistent regardless of temperature. The dynamics of dislocations in polycrystals were found to be independent of stress level and temperature. The Material Point Method (MPM) was used to explore snow avalanche behavior and identify flow regimes. The MPM accurately represented various flow patterns of snow avalanches, although challenges remained in capturing powder clouds. SVMs were employed for avalanche forecasting, using meteorological and snowpack variables as input features. The selected features provided insights into snowfall characteristics, snow accumulation, rain interaction, snowdrift patterns, cloud dynamics, snowpack mechanics, and temperature distribution within the snowpack. The findings contribute to a better understanding of avalanche dynamics and offer potential improvements in avalanche trend predictions.

## 1. Introduction

*1.1 Temperature variance*

Dislocations behave differently depending on the temperature. In fact, the mobility of a single basal dislocation in ice rises by a factor of around 15 from 20 to 3 °C. An increase in temperature may also encourage vacancy diffusion within the crystal, which will cause dislocations to climb. The purpose of this study was to ascertain whether temperature can affect how collective dislocation dynamics self-organize.

*1.2 Grain size variation*

We study the variation of certain aspects of an avalanche flow with respect to the particle sizes and other dynamics. In this case we have taken plastic avalanches as the point of study which is a phenomenon of cascading layers of degrading plastic because of various reasons. During the plastic avalanche, the deformation spreads rapidly and non-linearly throughout the material, involving the collective motion and interaction of numerous dislocations. The movement is usually taken over a metal surface. There is substantial evidence suggesting that in materials characterized by high dislocation mobility, such as face-centered cubic metals, the motion of dislocations occurs in a highly intermittent manner. This implies that plastic deformation occurs through isolated bursts or episodes, where the instantaneous strain rate can greatly exceed the average strain rate by several orders of magnitude. These bursts are temporary and localized events known as dislocation avalanches. There are multiple methods to study avalanches including (1) Studying the influence on physical properties that are sensitive to the density and velocity of mobile dislocations (2) Reading the electrical response of the stress recorded during and imposed strain rate test (3) Acoustic Emission

*1.3 Flow Regimes*

Snow exhibits different behaviors depending on the conditions, either resembling a fluid or a solid, which results in distinct characteristics of snow avalanches in reality. In the past, snow avalanches were typically categorized into two flow regimes: dense snow avalanches and powder snow avalanches. However, a recent study conducted by Köhler et al. (2018) shed light on the influence of snow temperature in classifying these flow regimes.
Here, we try to apply MPM (Material Point Method - a hybrid Eulerian–Lagrangian approach to solve and update the motion of the particles) in 2D (slope parallel and slope normal) to explore the distinct behaviors of snow avalanches and the key controlling factors of snow avalanche dynamics.

*1.4 SVMs*

We use Support Vector Machines (SVMs) in avalanche forecasting. The SVM algorithm is a popular supervised learning method used for classification problems. The goal of SVM is to create an optimal decision boundary, or hyperplane, that can accurately classify data points into different classes.

---

## 2. Method

*2.1 Temperature variance*

In this experiment, both cylindrical ice single crystals and polycrystals were prepared using distilled, deionized, and degassed water. The samples were cut at an angle to ensure that the preferred slip planes, known as basal planes, were inclined at the beginning of each test. The average grain size, denoted as $<d>$, ranged from 260 μm to 5 mm, which is relatively large compared to typical grain sizes in structural materials.

To achieve smaller grain sizes, fine-grained ice powders were pressed into dense aggregates under a low pressure of 0.1 Torr. Further, applying an axial stress of 9 MPa for a period of 2 hours resulted in a very small grain size of 260 μm, which is significant for ice. The c-axis orientations of the grains were not controlled, leading to an isotropic macroscopic mechanical behavior.

One advantage of using ice in this experiment is that both single crystals and polycrystals with different microstructures can be easily grown in the laboratory. Additionally, the transparency of ice allowed verification that the acoustic emission (AE) activity recorded was not related to microcracking. The ice samples were coupled with the AE transducer by melting/freezing. Parameters such as arrival time, maximum amplitude, and avalanche duration were recorded for each detected event. The system used two time constants to identify individual acoustic events and prevent the inclusion of secondary echoes. The durations of events depended on the chosen time constant and were compared between tests to analyze the damping of avalanches.

The experimental device was calibrated using a Nielsen test, which is a standardized artificial acoustic source.

*2.2 Grain size variation*

The method of experimentation used is the same as 2.1. The results obtained are done through different measurements and analysis.

*2.3 Flow Regimes*

The MPM method uses Lagrangian particles, assuming a continuous material, to track the properties such as mass, momentum, and deformation gradient. It also utilizes Eulerian grids to calculate the particle motion and update their states. The particle motion is determined by the conservation of mass and momentum, which can be expressed as follows:

$$\frac{D\rho}{Dt} + \rho \nabla \cdot \nu = 0$$
$$\rho \frac{D\nu}{Dt} = \nabla \cdot \sigma + \rho g$$

where $\rho$ is density, t is time, v is velocity, $\sigma$ is the Cauchy stress and g is the gravitational acceleration. Each particle in the MPM has a fixed mass, which naturally guarantees mass conservation.

To study snow avalanche behavior in controlled conditions, a setup featuring a rectangular snow sample and an ideal slope is used, as shown in Figure:

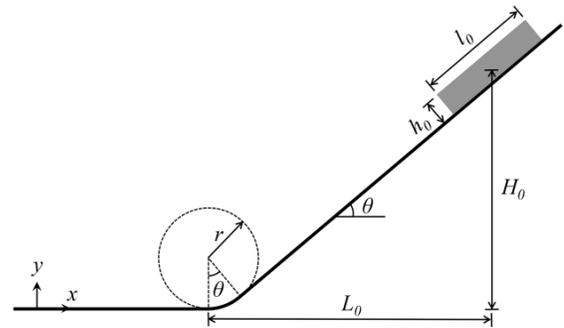

**Figure 1.** Model setup for MPM modeling of snow avalanches on an ideal slope.

The inclined slope is linked to the horizontal ground through a circular arc that has a central angle matching the slope angle θ. The circular arc and the horizontal ground are referred to as the connecting-arc zone and deposition zone, respectively.

In order to explore various flow patterns of snow avalanches, the characteristics such as the friction coefficient (M), the tension/compression ratio (β), the hardening factor (ξ), and the initial consolidation pressure are modified.

The impact of slope angle θ and horizontal length L₀ in Figure 1 is investigated through five sets of simulations outlined in Table 1. Each group examines different ranges of snow properties. Groups I, II, and III focus on studying the influence of slope angle $\theta$, while groups II, IV, and V analyze the effect of horizontal length $L_0$. In groups I, II, and III, the horizontal length $L_0$ remains fixed, and the drop height $H_0$ is adjusted as specified in Table 1 when varying $\theta$. Alternatively, one could fix the drop height $H_0$ and modify the horizontal length $L_0$. It is important to note that the ratios $L_0/h_0$, $L_0/l_0$, and $L_0/r$ are maintained constant when varying $L_0$ in groups II, IV, and V by adjusting $h_0$, $l_0$, and $r$ accordingly. Increasing $L_0$ results in scaling up the setup, which leads to an increase in the drop height $H_0$.

Table 1. Parameters adopted in the MPM simulations of snow avalanches on ideal slopes.

| | | Group I | Group II | Group III | Group IV | Group V |
|---|---|---|---|---|---|---|
| Snow | Density $\rho$ (kg m⁻³) | 250 | 250 | 250 | 250 | 250 |
| | Young's modulus $E$ (MPa) | 3 | 3 | 3 | 3 | 3 |
| | Poisson's ratio $\nu$ | 0.3 | 0.3 | 0.3 | 0.3 | 0.3 |
| | Friction coefficient $M$* | 0.1–1.5 | 0.1–1.5 | 0.1–1.5 | 0.1–1.5 | 0.1–1.5 |
| | Tension / compression ratio $\beta$* | 0.0–1 | 0.0–1 | 0.0–1 | 0.0–1 | 0.0–1 |
| | Hardening factor $\xi$* | 0.1–10 | 0.1–10 | 0.1–10 | 0.1–10 | 0.1–10 |
| | Initial consolidation pressure $p_0^{ini}$ (kPa)* | 3–30 | 3–30 | 3–30 | 3–30 | 3–30 |
| | Initial height $h_0$ (m) | 2 | 2 | 2 | 5 | 8 |
| | Initial length $l_0$ (m) | 12 | 12 | 12 | 30 | 48 |
| Slope | Bed friction coefficient $\mu$ | 0.5 | 0.5 | 0.5 | 0.5 | 0.5 |
| | Slope angle $\theta$ (°) | 30 | 40 | 50 | 40 | 40 |
| | Radius $r$ (m) | 10 | 10 | 10 | 25 | 40 |
| | Drop height $H_0$ (m) | 37.1 | 52.8 | 73.5 | 132.0 | 211.2 |
| | Horizontal length $L_0$ (m) | 65 | 65 | 65 | 162.5 | 260 |
| Simulation control | Mesh size (m) | 0.05 | 0.05 | 0.05 | 0.05 | 0.05 |
| | Time step (s) | 2.3 × 10⁻⁴ | 2.3 × 10⁻⁴ | 2.3 × 10⁻⁴ | 2.3 × 10⁻⁴ | 2.3 × 10⁻⁴ |
| | Frame rate (FPS) | 24 | 24 | 24 | 24 | 24 |

* $M$ values include 0.1, 0.5, 1.0 and 1.5. $\beta$ values include 0.0, 0.3, 0.6 and 1.0. $\xi$ values include 0.1, 0.5, 1.0 and 10.0. $p_0^{ini}$ values include 3, 12, 21 and 30 kPa.

The size of the background Eulerian mesh in the Material Point Method (MPM) is chosen to balance simulation accuracy and resolution, while also minimizing computation time. The time step is determined based on the CFL condition and the elastic wave speed to ensure simulation stability. The simulation data is exported at intervals of $1/24$ seconds. In each group of our MPM simulations, four typical flow regimes are captured with the changing mechanical properties of snow:

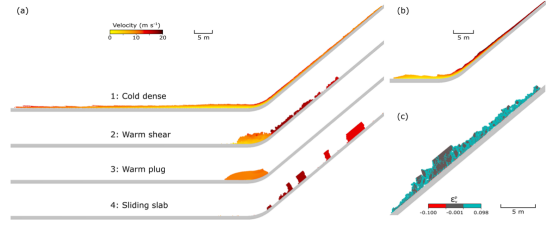

Figure 2. (a) Flows in four typical flow regimes captured at $t = 8.3$ s in the MPM simulations (Table 1, Group II). From top to bottom: cold dense, warm shear, warm plug and sliding slab. The cold dense flow is scaled up to be 3 times higher along the bed-normal direction for better visualization. (b) A flow with surges and small granules in transition from cold dense regime to warm shear regime at $t = 8.3$ s. The color denotes velocity as in (a). (c) The early stage ($t = 5.5$ s) of the warm shear flow in (a). Videos of the simulations are provided in the Supplement.

In the five groups of MPM simulations, where the slope angle $\theta$ and horizontal length $L_0$ are varied, a consistent pattern emerges between $v_{max}$ (maximum velocity) and α (slope angle): there is an initial decrease in $v_{max}$ followed by a constant $v_{max}$ as $\alpha$ increases. This two-stage relationship is evident in Figures 8 and 9.

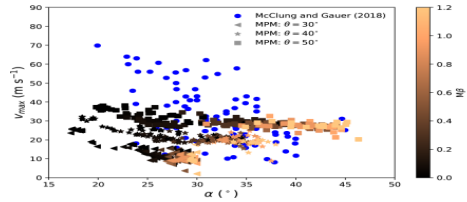

Figure 8. Evolution of the maximum velocity with α for varying slope angles $\theta$.

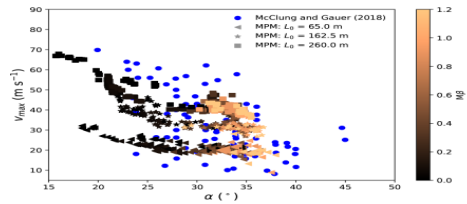

Figure 9. Evolution of the maximum velocity with α for different horizontal lengths $L_0$.

The first stage primarily includes cases with low friction and cohesion, while the second stage predominantly consists of cases with high friction and cohesion.

*2.4 SVMs*

The data used in this study consisted of daily measurements of meteorological and snowpack variables. To create an input feature vector, data from the current day and the two previous days were combined, resulting in a 30-dimensional vector. Expert features, including cumulative snow index, temperature changes, snow temperature gradients, and binary indicators for weather conditions and avalanche activity, were incorporated into the feature vector. Recursive feature elimination in conjunction with SVM was employed for feature selection, resulting in a final feature vector of 44 variables.

### 3. Review of literature

*3.1 Temperature and grain size variance*

In solid materials, acoustic emission (AE) waves are generated by sudden changes in inelastic strain. These waves can originate from different sources, including crack nucleation, crack propagation, twinning, or dislocation motion. However, in the specific study being discussed, the focus is on dislocation motion, as twinning does not occur in ice and experiments confirm the absence of crack nucleation.

The AE events detected in the study are most likely associated with synchronized and rapid motion of dislocations, known as dislocation avalanches, rather than isolated movements of individual dislocations. To understand the underlying source mechanism of these AE waves, a model is needed. The amplitude of the AE wave, represented by $A$, is related to the total length of dislocations involved in the plastic instability ($L$) and the average velocity of dislocations ($\nu$), as determined by theoretical analysis.

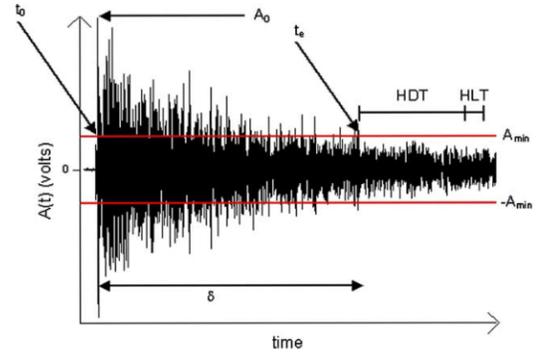

The formula that we get is

$$A(t) = \frac{3C_T^2}{4\pi C_L}\frac{\psi b}{D^2}L\nu = \frac{3C_T^2}{4\pi C_L}\frac{\psi b}{D^2}(\frac{dS}{dt})$$

where $C_T$ and $C_L$ are, respectively, the transverse and longitudinal wave velocities, $\phi$ is the density of the mate-rial and $D$ is the source/transducer distance (supposed to be large compared to $L$: far-field assumption. The term $1/D^2$ represents the geometrical attenuation of the acoustic waves. The term $L\nu$ accounts for the surface $S$ swept by time unit by dislocations during the avalanche: $L\nu = dS/dt$. When multiplied by $b$ and nor-malized by a volume, this term represents a strain rate $d\epsilon/dt$

*3.2 Flow regimes*

This study aims to showcase the effectiveness of the Material Point Method (MPM) in accurately representing various flow patterns of snow avalanches. The data from five reported real avalanches with different complex terrains are analyzed and contrasted with their respective MPM models. The evolution of the scaled front velocity is examined along the flow path. The field data for the front velocity was collected using Doppler radar devices and photo analyses.

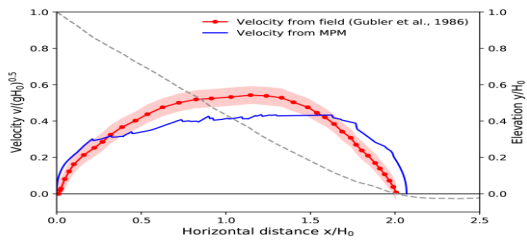

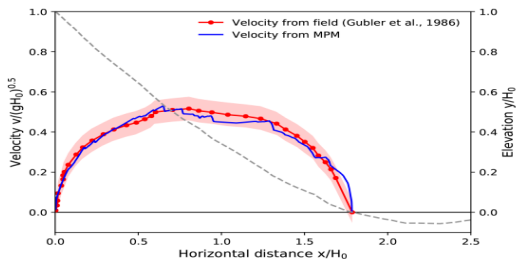

**Figure 11.** Front velocity distribution along the flow path for Case I: Weissfluh north ridge, 12 March 1982, a1 (Davos, Switzerland). Drop height $H_0 = 236$ m.

**Figure 12.** Front velocity distribution along the flow path for Case II: Weissfluh north ridge, 12 March 1982, a2 (Davos, Switzerland). Drop height $H_0 = 177$ m.

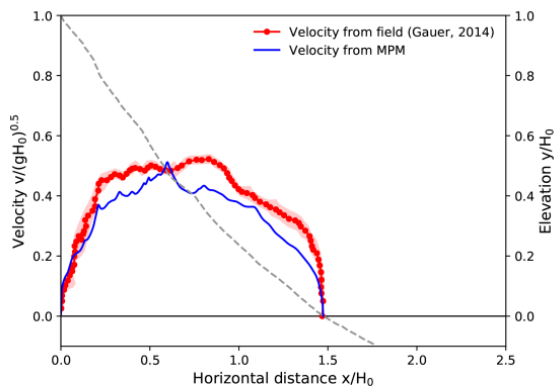

**Figure 13.** Front velocity distribution along the flow path for Case III: Himmelegg, 14 February 1990 (Arlberg, Austria). Drop height $H_0 = 352$ m.

The consistency between the MPM results and the measured data is better for the second avalanche in Fig. 12. The underestimated maximum front velocity in Fig. 11 might be due to the challenge of capturing the powder cloud of the first avalanche with the MPM. the neglection of entrainment in the simplified MPM simulation may also contribute to the discrepancy in Fig. 11

*3.3 SVMs*

The selected features in the study have direct relevance to Fluid Particle Mechanics (FPM) in the context of snow and precipitation. The new-snow index, cumulative snow index, rain at 900 m, snow drifting, cloud cover, foot penetration, and snow temperature all have connections to FPM. These features provide insights into snowfall characteristics, snow accumulation, rain interaction, snowdrift patterns, cloud dynamics, mechanical properties of the snowpack, and temperature distribution within the snowpack.

## 4. Discussion

*4.1 Temperature variance*

The study demonstrates that the distribution of avalanche sizes in ice single crystals remains unchanged regardless of applied shear stress and temperature. This highlights the robustness of the scale-free pattern, which arises from long-range elastic interactions between dislocations and is independent of individual dislocation behavior. Lattice friction, a thermally activated process, has minimal influence on the scale-free critical dynamics even at temperatures close to the melting point. The proportion of decorrelated dislocations and drag resistance from phonon interactions become significant factors at high velocities, with the latter increasing with temperature. Additionally, strain hardening and temperature affect the durations of dislocation avalanches, with higher temperatures leading to shorter durations and faster dislocation velocity decay.

In conclusion, the study confirms the temperature independence and universality of the scale-free critical dynamics in ice. The interplay of temperature, lattice friction, phonon drag resistance, and strain hardening contributes to the dynamics of dislocation avalanches. Further investigations, particularly through numerical simulations, can offer deeper insights into these mechanisms and their implications for plastic deformation in ice.

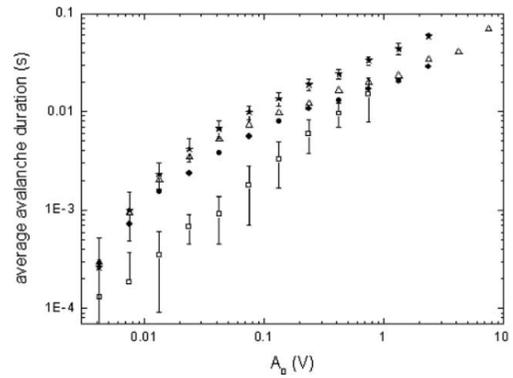

Fig. 8. Avalanche durations in polycrystals. Comparison of the average duration of events of amplitude $A_0$ between a single crystal and different polycrystals, as a function of $A_0$. Open squares: $\langle d \rangle = 0.26$ mm; applied compression stress: 0.67 MPa. Closed circles: $\langle d \rangle = 0.87$ mm; applied compression stress: 0.54 MPa. Open triangles: $\langle d \rangle = 5.02$ mm; applied compression stress: 0.55 MPa. Closed stars: single crystal; applied compression stress: 0.40 MPa. All the test were performed at $-10$ °C. We checked that, for a given $A_0$, the duration data obey the central limit theorem. Therefore, averaging the durations is sensible. For a question of graph readability, error bars are only represented for the single crystal and the 0.26 mm grain size polycrystal. Otherwise error bars would overlap.

### 4.2 Grain size variation

The dynamics of dislocations in polycrystals were found to be independent of stress level and temperature, similar to single crystals. Monte-Carlo simulations were conducted to validate the estimation of the power law exponent. The analysis of the 3D spatial organization of dislocation avalanches using correlation integral analysis revealed that avalanches exhibited spatial correlation over distances one order of magnitude larger than the average grain size. From a macroscopic perspective, the slowing down of dislocation motion during creep tests indicated hardening of the sample. Hardening can result from the development of long-range internal stresses (kinematic hardening) and/or short-range dislocation interactions (isotropic hardening). Comparatively, the average grain size was smaller in polycrystals than in single crystals, and it decreased with decreasing grain size. This suggests a relationship between strain hardening and the damping of avalanches, with larger strain hardening leading to quicker damping. Notably, for avalanches larger than 0.01 V, the duration (d) was consistently longer during unloading than during loading.

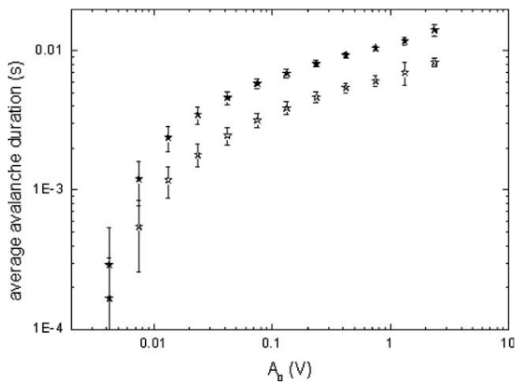

Fig. 4. Evolution of avalanche durations with temperature in single crystals. Comparison of the average duration of events of amplitude $A_0$ between single crystals tested at different temperatures, as a function of $A_0$. Closed symbols: $T = -10$ °C; resolved shear stress: 41 kPa. Open symbols: $T = -3$ °C; resolved shear stress: 67 kPa.

*4.3 Flow regimes*

The broad division of Avalanche flows into four regimes could accommodate the different cases that were considered. The evolution of the avalanche front, the shape of the free surface, and the vertical velocity profile exhibit unique characteristics for each flow regime, playing a crucial role in identifying and distinguishing between different flow regimes. The behavior of dense snow avalanches has been well recovered by the MPM. A discrepancy was observed particularly for avalanches which developed a powder cloud above the dense core, as the powder cloud has not been modeled here.

*5.4 SVMs*

The findings suggest that the selected features provide valuable insights into fluid dynamics and particle mechanics involved in snowfall, snowpack formation, and related processes. By incorporating FPM concepts, such as snowflake characteristics, snow accumulation effects, rain interaction, snowdrift patterns, cloud dynamics, snowpack mechanics, and thermal properties, the SVM-based avalanche forecasting model can leverage inherent FPM features to improve avalanche trend predictions.

## 5. Result

*5.1 Temperature variance*

Ice single crystals were subjected to uniaxial compression creep experiments at temperatures of 3°C, 10°C, and 20°C. Three different methods were used to set the compression stress as a function of temperature: maintaining the resolved shear stress on basal planes, maintaining the macroscopic strain rate, and maintaining the average velocity of a single dislocation. The distribution of avalanche sizes was found to be independent of the applied shear stress and unaffected by temperature. The avalanche size distributions collapsed precisely after being renormalized by the total number of avalanches, following the same power law. This indicates that the scale-free pattern of avalanche sizes is independent of temperature. Furthermore, the scale-free critical dynamics were shown to be independent of individual dislocation behavior and instead result from long-range elastic interactions between dislocations.

*5.2 Grain size variance*

In polycrystals, the average grain size plays a significant role due to the interaction between dislocations and grain boundaries (GBs). GBs act as barriers to dislocation motion, leading to the formation of dislocation pile-ups and internal stresses. They can activate neighboring dislocation sources and allow dislocation transmission in specific cases. Additionally, GBs can serve as effective sources of dislocations.

The presence of GBs as barriers may impact the large-scale propagation of dislocation avalanches and modify the scale-free dynamics observed in single crystals. To investigate this, acoustic emission (AE) was recorded during plastic deformation of ice polycrystals with different average grain sizes, and the statistical data were compared to single crystal results. Creep tests revealed a decrease in acoustic activity as the material hardened, indicating a decrease in the visco-plastic strain rate. Stationary creep showed minimal AE activity, suggesting that processes such as dislocation climb in tilt boundaries may not contribute to AE. However, transient creep, associated with basal slip in ice polycrystals, was a significant source of AE. The distribution of avalanche sizes in polycrystals differed from the scaling observed in single crystals, with a smaller power law exponent and a cut-off for large amplitudes.

*5.3 Flow regimes*

The calculated avalanche front position and velocity from the MPM show reasonable agreement with the measurement data from the literature.

*5.4 SVMs*

The SVM classifier trained on the selected features demonstrated the ability to capture important information related to snowpack dynamics. The chosen features retained significant Class II (snowpack) information, including subjective foot-penetration values. Notably, expert features related to wind conditions, particularly south or southeasterly winds, were emphasized. Only two non-expert features from two days prior to the forecast day, foot penetration and wind direction, were retained, indicating the rapid changes in Scotland's maritime climate.

---

6. **Conclusion**

*6.1 Temperature variance*

The scale-free pattern observed in single crystals was found to be unaffected by temperature. However, temperature did have an impact on the durations of avalanches, with higher temperatures resulting in shorter durations. This observation was attributed to interactions between dislocations and phonons. Furthermore, the influence of microstructure was examined by conducting tests on polycrystals with varying average grain sizes.

*6.2 Grain size variance*

The introduction of the average grain size as a microstructural scale in polycrystalline ice disrupted the scale-free pattern observed in single crystals. Grain boundaries (GBs) were found to play a dual role as barriers to the dynamic propagation of dislocation avalanches and as transmitters of internal stresses. This suggests a complex effect of GBs on the mechanical behavior of the material. The relaxation of dislocation avalanches was influenced by the presence of GBs, as indicated by the decrease in avalanche durations with increasing levels of hardening. Additionally, dislocation avalanches were observed during reverse dislocation motion in polycrystalline ice, and the power law exponent for these avalanches was consistent with the scaling observed in single crystals. These findings highlight the importance of considering microstructural factors, such as grain size and GBs, in understanding the dynamics of dislocation avalanches and the mechanical behavior of polycrystalline materials.

*6.3 Flow regimes*

Despite the simplifications and assumptions made in this study, the findings demonstrate that the MPM model holds great potential for conducting systematic and quantitative investigations into the dynamics of snow avalanches and the transitions between different flow regimes. This approach allows for a comprehensive analysis of the interplay between snow mechanical properties and terrain geometries. Such research contributes to a better understanding of wet snow avalanches and provides insights into analyzing avalanche dynamics under the influence of climate change.

*6.4 SVMs*

The study demonstrates the effectiveness of SVMs and expert-selected features in avalanche forecasting. The combination of meteorological and snowpack variables, along with additional expert features, provides valuable information for understanding fluid dynamics and particle mechanics in snow-related processes. Further research can explore the integration of FPM concepts into avalanche forecasting models to enhance prediction accuracy and contribute to improved safety measures.

---

**Appendix**

**A-1: Neilsen Test**
The Hsu-Nielsen source, also known as pencil lead break, is a device used to simulate an acoustic emission event. It involves breaking a brittle graphite lead of 0.5 mm (or alternatively 0.3 mm) diameter, located within a suitable fitting, approximately 3 mm (+/- 0.5 mm) from its tip by pressing it against the surface of the object under examination. This action generates a strong acoustic signal, similar to a natural acoustic emission source, which can be detected by sensors as a burst of activity.

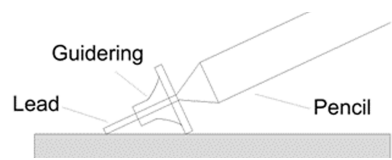

**A-2: Recursive feature elimination**
RFE is a feature selection technique that gradually eliminates the weakest feature(s) from a model until a specified number of features is reached. RFE combines cross-validation with feature scoring to evaluate different subsets of features and select the best-performing set. The RFECV visualizer provides a visual representation of the number of features in the model, their cross-validated test scores, and variability, allowing for the visualization of the selected number of features.

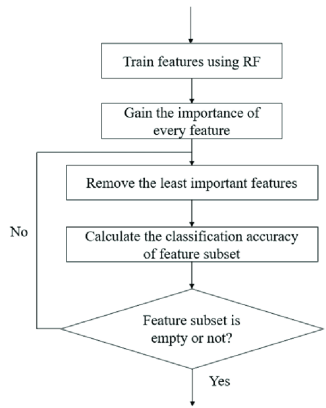

**A-3: Powder Snow Avalanches**
Characterized by the suspension and movement of snow grains in a state of fluid turbulence, and primarily driven by the action of gravity, these avalanches resemble particle-laden gravity currents and exhibit similarities to other natural phenomena such as turbidity currents, pyroclastic flows, and desert dust storms.

---

## References

- *Dislocation avalanches: Role of temperature, grain size and strain hardening* [Thiebaud Richeton *et al.*]
- *Applying machine learning methods to avalanche forecasting* [A. Pozdnoukhov *et al.*]
- *The mechanical origin of snow avalanche dynamics and flow regime transitions* [Xingyue Li *et al.*]
- *Snow Avalanches* [C. Ancey]